\documentclass[aps,prl,twocolumn,superscriptaddress]{revtex4-2}

\usepackage{amsmath,amssymb,amsfonts}
\usepackage{bm}
\usepackage{booktabs}
\usepackage{microtype}
\usepackage{hyperref}

\hypersetup{
  colorlinks=true,
  linkcolor=blue,
  citecolor=blue,
  urlcolor=blue
}

\begin{document}

\title{Small-System Group: Thermodynamics as a Complete Self-Similarity Limit}

\author{Amilcare Porporato}
\affiliation{Princeton University}
\email{aporpora@princeton.edu}

\author{Lamberto Rondoni}
\affiliation{Department of Mathematical Sciences, Politecnico di Torino, Turin, Italy}
\affiliation{INFN, Sezione di Torino, Turin, Italy}
\email{lamberto.rondoni@polito.it}

\date{\today}

\begin{abstract}
We revisit the Rayleigh--Riabouchinsky paradox in dimensional analysis by making the bridge between thermodynamics and the mechanical description of temperature explicit. Introducing Boltzmann's constant $k_B$ as a dimensional unifier leads to an augmented $\Pi$-theorem with an additional dimensionless group related to system size. In the macroscopic thermodynamic limit, this small-system group, $\Pi_B = k_B/(c\,\ell^3)$---the inverse heat capacity of a control volume of size $\ell^3$ in units of $k_B$---becomes irrelevant as the response becomes self-similar with respect to it, recovering Rayleigh's formulation. Our results recast thermodynamics as the complete-similarity limit of statistical mechanics with respect to the small-system group, a parameter that also controls thermodynamic fluctuations. We also discuss second-order phase transitions from the viewpoint of incomplete similarity.
\end{abstract}

\begin{abstract}
We revisit the Rayleigh--Riabouchinsky paradox in dimensional analysis by making explicit the bridge between thermodynamics and the mechanical interpretation of temperature. Boltzmann's constant $k_B$ acts as a dimensional unifier, leading to an augmented $\Pi$-theorem with an additional dimensionless group that encodes system size. In the macroscopic thermodynamic limit this small-system group,
$\Pi_B = k_B/(c\,\ell^3)$---the inverse heat capacity of a control volume of size $\ell^3$ in units of $k_B$---becomes irrelevant as the response becomes self-similar with respect to it, recovering Rayleigh's formulation. Under suitable conditions, macroscopic limits make the fluctuations of the observables of interest negligible compared to their expected values, hence the state of a system is characterized by a reduced set of parameters. We thus recast thermodynamics as the complete-similarity limit of statistical mechanics with respect to $\Pi_B$, which also controls thermodynamic fluctuations. We also discuss second-order phase transitions from the viewpoint of incomplete similarity.
\end{abstract}

\maketitle

In 1915, a dispute arose over what dimensional analysis should predict for a forced-convection heat-transfer problem. Rayleigh’s thermodynamic formulation \citep{Rayleigh1915a} treated temperature as a primary dimension and implied a single governing dimensionless group, whereas Riabouchinsky \citep{Riabouchinsky1915} argued that a purely mechanical description should suffice since temperature is reducible to energy; yet doing so introduced an additional similarity parameter, which seemed paradoxical.

Working in the thermodynamic dimensional basis $\{L,T,M,\Theta\}$ (length, time, mass, temperature) \citep{Barenblatt1996}, Rayleigh considered a body of characteristic size $\ell$ in a uniform stream of incompressible, inviscid fluid of speed $v$. The body transfers heat to the fluid at a characteristic rate $h$, driven by a temperature difference $\theta$. The relevant material parameters are the volumetric heat capacity $c$ and thermal conductivity $\kappa$, with dimensions
$[\ell]=L$, $[v]=L\,T^{-1}$, $[h]=M\,L^{2}\,T^{-3}$, $[\theta]=\Theta$,
$[c]=M\,L^{-1}\,T^{-2}\,\Theta^{-1}$, and $[\kappa]=M\,L\,T^{-3}\,\Theta^{-1}$.
The corresponding law \citep{Porporato2022} may be written as $h=f(\ell,\theta,c,\kappa,v)$.
Treating temperature as an independent dimension, the variables $(h,\ell,\theta,c,\kappa,v)$ yield a single nontrivial similarity relation by the Buckingham--$\Pi$ theorem \citep{Barenblatt1996,Bridgman1922}:
\begin{equation}
  \frac{h}{\ell\,\theta\,\kappa} \;=\; \phi\!\left(\frac{v\,c\,\ell}{\kappa}\right),
  \label{eq:Rayleigh}
\end{equation}
where the argument is the (thermal) P\'eclet number (the ratio of diffusive to advective timescales).

Riabouchinsky objected that statistical mechanics identifies temperature with mean molecular kinetic energy, so temperature should not be treated as dimensionally independent of energy \citep{Riabouchinsky1915}. If temperature is eliminated as a fundamental dimension---working, for example, with only length, time, and energy---the same variable list implies an additional independent dimensionless group, contradicting Eq.~\eqref{eq:Rayleigh}.  Rayleigh acknowledged that the issue was ``worthy of further discussion'' while arguing that Fourier's heat equation effectively treats heat and temperature as \emph{sui generis}, legitimizing his thermodynamic choice of dimensions \citep{Rayleigh1915b}. Buckingham called for clarification \citep{Buckingham1915}, but the paradox continued to be source of investigation up to this day (e.g., \citep{Gibbing19080,West1984,Butterfield2001,Porporato2022}). 

Revisiting the controversy is timely because it highlights issues that reappear in small-system thermodynamics, an area still seeking fully consistent formulations. Modern nanotechnologies and single-particle experiments probe regimes where fluctuations are comparable to mean values and finite-size effects blur distinctions that are sharp in classical thermodynamics. In the microscopic mechanical description, energy can only be seen in two forms: kinetic and potential, which would justify Riabouchinsky's point of view once $\theta$ is identified with the second. 
On the other hand, for large numbers of suitably interacting particles,  
one further distinction can be made, the one between the 
statistical notions of order and disorder, 
and that makes such numbers crucial.  
While this distinction is sharp for thermodynamics of macroscopic objects, it makes no sense for a few particles. For intermediate domains, the modern science and technology of small systems has revitalized attention to the thermodynamics of fluctuations and motivated newer frameworks such as stochastic thermodynamics and nanothermodynamics
\citep{ECM93,Jarzynski1997,Crooks1999,Seifert2012,Ciliberto2017,PorporatoCalabreseRondoni2024TLE}. In this setting, questions that are often negligible macroscopically become central: when is temperature operationally well defined, how do finite reservoirs modify canonical predictions, and what controls the approach to the thermodynamic limit?

Dimensional analysis suggests that when a governing dimensionless group becomes very small or very large, the response may approach a self-similar asymptotic regime: the group can drop out (complete similarity) or persist only through nontrivial exponents (incomplete similarity) \citep{Barenblatt1996}. In the present setting, we show that incorporating microscopic or finite-reservoir structure introduces an additional size-dependent parameter, and that the macroscopic theory is recovered only as this parameter is driven to its asymptotic limit. As a result, a natural measure of thermodynamic system size emerges as a dimensionless heat capacity, which controls both finite-size corrections and thermometric uncertainty; the thermodynamic limit corresponds to this dimensionless group tending to zero, yielding a complete self-similarity limit. Note that this limit is not typically  obtained analytically but via calculations or experimentally \citep{Barenblatt1996}.

{\it Augmented Dimensional Analysis.} For a law relating $n$ variables, dimensional analysis enforces invariance under changes of units by requiring dimensional homogeneity. In a chosen basis of fundamental dimensions one forms the dimensional matrix $\bm A$ of exponents, whose rank $r=\mathrm{rank}(\bm A)$ counts the independent dimensional constraints \citep{Barenblatt1996,Bridgman1922}. The Buckingham--$\Pi$ theorem then implies that the law can be rewritten as a relation among $n-r$ independent dimensionless groups, constructed (nonuniquely) by choosing $r$ repeating variables and eliminating dimensions \citep{Bridgman1922,Barenblatt1996}. In practice, selecting a dimensional basis encodes modeling judgment \citep{Porporato2022} and when one changes the dimensional basis by identifying previously independent dimensions (here, relating energy and temperature), the transformation must be mediated by a dimensional unifier \citep{Panton2006,Porporato2022}.

Thus, unifying Rayleigh's and Riabouchinsky's analyses requires an explicit conversion factor linking energy and temperature. The  Boltzmann's constant, $k_B=1.38 \cdot 10^{-23}$ J/K, a unit of entropy, plays that role. With it impressively low numerical value it  measures the ``distance''
between the microscopic world and the macroscopic world, bridging mechanics and thermodynamics. From a dimensional point of view, this leads to the augmented variable list
\begin{equation}
  h \;=\; f_a(\ell,\theta,c,\kappa;\,v,k_B),
  \label{eq:augmentedlaw}
\end{equation}
from which, with $(\ell,\theta,c,\kappa)$ as repeating variables, one obtains the two-parameter similarity form
\begin{equation}
  \frac{h}{\ell\,\theta\,\kappa}
  \;=\;
  \phi_a\!\left(
    \frac{v\,c\,\ell}{\kappa},
    \frac{k_B}{c\,\ell^3}
  \right).
  \label{eq:augmentedPi}
\end{equation}
The second argument,
\begin{equation}
  \Pi_B \;=\; \frac{k_B}{c\,\ell^3},
  \label{eq:boltzmanngroup}
\end{equation}
compares a molecular thermal scale set by $k_B$ to the macroscopic heat capacity of a volume $\ell^3$. 

{\it Small-System Group.} Defining the heat capacity of the control volume as $C_\ell \equiv c\,\ell^3$, one may write
\begin{equation}
\Pi_B=\frac{k_B}{C_\ell},
\end{equation}
so $\Pi_B$ is the inverse heat capacity measured in units of $k_B$. This interpretation connects directly to fluctuation theory. For a finite system in thermal equilibrium, standard arguments \cite{Callen1985} give relative (kinetic) temperature uncertainty scaling like $\sqrt{k_B/C_\ell}$, so $\Pi_B$ controls the degree of self-averaging of temperature at the scale $\ell$ \citep{Falcioni2011}. 
$\Pi_B\ll 1$ implies the system is very large, which, together with a sufficient degree of disorder, makes 
microscopic fluctuations negligible, temperature sharply defined, and normal thermodynamics applicable. 
When $\Pi_B$ is not small, thermal fluctuations and finite-size effects become relevant and the system must be considered thermodynamically small. 
In many microscopic models one expects $C_\ell \sim N_\ell k_B$, where $N_\ell$ is the number of effective degrees of freedom contained in $\ell^3$, so $\Pi_B \sim 1/N_\ell$, making explicit that $\Pi_B$ plays the role of an inverse system-size parameter. Clearly, the maximum $N_\ell$ in a box of side $\ell$
is limited by the size or range of interaction of the particles constituting the system.

Thus, the role of the similarity group $\Pi_B$ extends well beyond Rayleigh’s convection problem. It is closely related to the parameters that control temperature uncertainty and finite-reservoir effects in statistical mechanics and in thermometry of small systems \citep{Callen1985,WuWidom1998,Falcioni2011,Richens2018}. In mesoscopic calorimetry and quantum heat-transport experiments the same control parameter appears explicitly as the dimensionless heat capacity $\eta\equiv C/k_B=1/\Pi_B$, which sets the fundamental scale of effective temperature fluctuations for a finite absorber coupled to a bath \citep{PekolaKarimi2021}.
 
For a perfect gas, the group $\Pi_B$ admits a particularly transparent interpretation. Let $n$ denote the number density (particles per unit volume), and let $f$ be the number of active quadratic degrees of freedom per particle (e.g.\ $f=3$ for a monatomic gas in the classical regime). Since the internal energy density is
\begin{equation}
u=\frac{f}{2}\,n k_B T,
\end{equation}
the volumetric heat capacity at constant volume is
\begin{equation}
c_v = \left(\frac{\partial u}{\partial T}\right)_V=\frac{f}{2}\,n k_B.
\end{equation}
Substituting into Eq.~\eqref{eq:boltzmanngroup} yields
\begin{equation}
\Pi_B=\frac{k_B}{c_v\,\ell^3}
=\frac{k_B}{\left(\frac{f}{2}\,n k_B\right)\ell^3}
=\frac{2}{f\,n\ell^3}
=\frac{2}{f\,N_\ell},
\label{eq:PiB_ideal_gas_cv}
\end{equation}
where $N_\ell=n\ell^3$ is the expected number of particles in the control volume. 

{\it Thermodynamic Limit as Complete Self-Similarity.} For macroscopic bodies,
$\Pi_B = k_B/(c\,\ell^3)$ is typically extremely small, becoming non-negligible only at
sufficiently small $\ell$ (e.g., nanoscale systems or ultra-small thermal masses). For
example, for water at room temperature $c \sim 4\times 10^{6}\ \mathrm{J\,m^{-3}\,K^{-1}}$, so for $\ell = 1\,\mathrm{mm}$, the control-volume heat capacity is
$C_\ell = c\,\ell^{3} \sim 4\ \mathrm{J\,K^{-1}}$, so $C_\ell/k_B \sim 3\times 10^{23}$
and hence $\Pi_B = k_B/C_\ell \sim 10^{-23}$.
The thermodynamic regime corresponds to $\phi_a(\cdot,\cdot)$ in Eq.~\eqref{eq:augmentedPi}
approaching an asymptotic plateau as $\Pi_B\to 0$. It is an empirical fact that the dependence on $\Pi_B$
disappears (complete self-similarity of the first kind) and
\begin{equation}
  \frac{h}{\ell\,\theta\,\kappa}
  \;=\;
  \lim_{\Pi_B\to 0}\phi_a\!\left(
    \frac{v\,c\,\ell}{\kappa},\Pi_B
  \right)
  \;=\;
  \phi\!\left(\frac{v\,c\,\ell}{\kappa}\right),
  \label{eq:plateau}
\end{equation}
recovering Rayleigh's single-group form \eqref{eq:Rayleigh}. The key point is that
Rayleigh's thermodynamic analysis effectively assumes the existence of this asymptotic
regime. The reduction is therefore an empirical statement about insensitivity to
molecular details in the macroscopic limit, not an automatic algebraic consequence of
$\Pi_B$ being small. In this limit, 
that includes sufficiently fast decay of correlations in space and time,
microscopic information is effectively compressed
into a low-dimensional macroscopic description (cf.\ \cite{Callen1985,Mandelbrot1962}).

{\it Regular and Singular Thermodynamic Limits.}---
When the limit $\Pi_B\to 0$ is regular, the response approaches a plateau and dependence
on $\Pi_B$ drops out, with finite-$\Pi_B$ corrections organized by $\Pi_B$. For temperature
estimation with a finite thermometer, estimation theory gives
\begin{equation}
\mathrm{Var}(\widehat{T}) \sim \frac{k_B T^2}{C},
\label{eq:T_estimator_variance}
\end{equation}
so the relative uncertainty scales like $\sqrt{\Pi_B}$ when $C\equiv C_\ell$
\citep{Falcioni2011}. Likewise, deriving the canonical distribution from a finite bath
produces corrections controlled by $k_B/C_B$: expanding the bath entropy yields
\begin{equation}
p(E)\propto \exp\!\left(-\beta E-\frac{E^2}{2\,C_B\,k_B\,T^2}+\cdots\right),
\label{eq:finite_bath_expansion}
\end{equation}
so deviations from the Boltzmann factor are organized by the bath analogue of $\Pi_B$
\citep{Richens2018,Ramshaw2018,Biro2022,PlastinoPlastinoRocca2023}. Exactly
solvable finite-size models make the approach explicit: for the two-dimensional Ising
model on a torus, energy and specific heat at criticality admit controlled finite-size
expansions, and more generally finite-size scaling replaces the singular thermodynamic
limit by smooth crossover functions \citep{FisherBarber1972,Salas2001}.

The plateau associated with $\Pi_B\to 0$ can nevertheless fail in singular regimes,
notably at continuous phase transitions, where scale invariance emerges but dimensional
analysis does not fix exponents (Barenblatt's incomplete similarity of the second kind)
\citep{Barenblatt1996,Porporato2022, Goldenfeld1992}. Near criticality, the correlation length
diverges and observables follow power laws in the reduced temperature
$t=(T-T_c)/T_c$, with exponents set by dimension and symmetry (universality class)
\citep{WilsonKogut1974,Fisher1974,Goldenfeld1992}; the same applies dynamically,
where $\tau\sim \xi^{z}$ and $z$ depends on conservation laws and order-parameter dynamics
\citep{HohenbergHalperin1977,Taeuber2014}. Some two-dimensional systems instead exhibit an
essential singularity \citep{Kosterlitz1974}, and near criticality
finite size competes with $\xi$, leading to finite-size scaling forms
\citep{Privman1990}.

{\it Conclusion.}---The Rayleigh--Riabouchinsky controversy highlights how macroscopic theories are recovered only in an asymptotic regime in which an appropriate
small-system group becomes negligible. Here that group is the inverse heat capacity in
units of $k_B$, $\Pi_B=k_B/(c\,\ell^3)$, which quantifies how efficiently
thermal degrees of
freedom self-average within a control volume of size $\ell^3$, 
if the dynamics are 
sufficiently random.
In Barenblatt's terminology,
the plateau associated with the thermodynamic limit corresponds to complete
self-similarity \citep{Barenblatt1996}.

This viewpoint connects naturally to thermodynamic fluctuations and small-system
thermodynamics. In such regimes, $C/k_B$ (or its inverse $\Pi_B$) acts as a control
parameter organizing both estimation uncertainty and finite-size corrections. More broadly,
the same logic clarifies when asymptotic simplifications are regular and when they are not:
in regular limits one expects a plateau with corrections expandable in $\Pi_B$, whereas in
singular limits---notably near critical points, where correlations persist---the plateau can fail and incomplete
similarity emerges, with nontrivial exponents or crossover forms determined by internal
statistical-mechanical structure. More generally, long-range interactions, strong
correlations, or constrained phase-space growth can alter the scaling of $C_\ell$ with
$\ell$ and thereby modify the approach to the macroscopic regime \citep{Barenblatt1996}.
The size of $\ell$ for the thermodynamic fields to make sense (the condition of local thermodynamic equilibrium) depends on the range of interactions of the microscopic constituents of matter \cite{Kreuzer,Spohn2012,Falcioni2007,MMLR}.

\begin{acknowledgments}
\end{acknowledgments}

\bibliographystyle{apsrev4-2}

\end{document}